# Real space imaging of metastable Bragg glass states in a weakly pinned Type II superconductor

Garima Saraswat[a], Somesh Chandra Ganguli[a*], Rini Ganguly[a], Parasharam Shirage[b], Vivas Bagwe[a], Anand Kamlapure[a], Arumugam Thamizhavel[a] and Pratap Raychaudhuri[a†]

[a] Tata Institute of Fundamental Research, Homi Bhabha Road, Colaba, Mumbai 400005, India.
[b] Indian Institute of Technology Indore, IET-DAVV Campus, Khandwa Road, Indore 452017, India.

The breakdown of crystalline order in a disordered background connects to some of the most challenging problems in condensed matter physics. For a superconducting vortex lattice, the equilibrium state in the presence of impurities is predicted to be a "Bragg glass" (BG), where the local crystalline order is maintained everywhere and yet the global positional order decays algebraically. Here, using scanning tunnelling spectroscopy (STS) we image the vortex lattice in a weakly pinned $NbSe_2$ single crystal. We present direct evidence that the ordered state of the VL is a BG, consisting of a large number of degenerate metastable states, which is a hallmark of a glassy state. These results are a significant step towards understanding the disordering of a lattice under the influence of quenched random disorder with a direct impact on various fields, including charge density waves, colloidal crystals and self-organised periodic structures on a substrate.

---

* e-mail: someshcg@tifr.res.in
† e-mail: pratap@tifr.res.in

**I. Introduction**

The vortex lattice (VL) in a Type II superconductor provides an ideal system to study various notions related to disordering and melting[1,2,3]. In a clean superconductor, the VL configuration is determined by two factors: The interaction between vortices which favours their ordering into a hexagonal Abrikosov lattice, and temperature which favours disordering and melts the lattice at a characteristic temperature. Structural imperfections in the superconductor modify this simple scenario by creating a disordered background potential, where the vortices can get pinned at individual defect sites. In the presence of random pinning the VL can no longer support a true long-range order. Of particular interest is the VL state in the presence of weak random pinning well below the locus of the thermodynamic order-disorder transition in magnetic field-temperature (*H-T*) parameter space. It has been predicted that in equilibrium, the VL exists in a quasi-long-range ordered state, the so called Bragg glass (BG)[4,5], where each lattice point maintains six-fold coordination and yet the positional order decays algebraically with distance. The hallmark of such a state is the existence of a large number of degenerate metastable configurations, each with algebraic decay of positional order but different vortex configuration. With increase in temperature/magnetic field this quasi-long-range order is destroyed giving rise to a disordered state[6].

Both ordered and disordered VL states, as well as order-disorder transitions have been extensively studied through bulk measurements such as critical current[7,8], a.c. susceptibility[9,10,11] and d.c. magnetisation[12,13]. These studies rely on the fact that in the presence of weak random pinning sites, a disordered VL is more strongly pinned to the crystal than its more ordered counterpart[14]. However, a more direct and relatively less explored line of study has been through imaging of the VL in real space[15,16,17], which gives direct information on the lattice configuration. Here, using STS down to 350 mK, we image in real space equilibrium and non-equilibrium states of the VL in a weakly pinned NbSe$_2$ single crystal, well below the locus[18,19] of the thermodynamic order-disorder transition (see supplementary information[20]). We use thermo-magnetic cycling to prepare the VL in a variety of different states, e.g. non-equilibrium disordered states and equilibrium quasi-long range ordered states and use small magnetic pulses to switch between them. Our investigations reveal that while the VL can be prepared both as disordered and quasi-long-range ordered states, the equilibrium state of the system is a BG, composed of large number of degenerate metastable states, which can be accessed using small magnetic perturbations.

## II. Experimental Details

NbSe$_2$ single crystals were grown by iodine vapor transport method, starting with 99.99 % pure Nb powder and 99.999 % Se shots. Stoichiometric amounts of pure Nb and Se, together with iodine as the transport agent were mixed and placed in one end of a quartz tube, which was then evacuated and sealed. The sealed quartz tube was heated up in a temperature gradient furnace for 5 days, with the charge-zone and growth-zone temperatures, 800°C and 720°C respectively. We obtained NbSe$_2$ single crystals with lateral size 4-5 mm. Resistivity ($\rho$) measurements showed a superconducting transition temperature, $T_c$ ~ 7.2 K with transition width of $\Delta T_c$ ~ 0.15 K and a residual resistance ratio $\rho$ (300 K)/$\rho$ (10 K) ~ 25.

The VL is imaged using a home-built scanning tunneling microscope[21] (STM) operating down to 350 mK and fitted with an axial 9 T superconducting solenoid. Prior to STM measurements, the crystal is cleaved in-situ in vacuum giving atomically smooth facets larger than 1 μm × 1 μm. Since the vortex core behaves like a normal metal, well resolved images of the VL are obtained by measuring the tunneling conductance, *G (V) = dI/dV*, at a fixed bias voltage (*V*) on the surface of the superconductor. We fix *V* ~ 1.4 mV, which is close to the superconducting coherence peaks such that each vortex core manifests as a local minimum in *G (V)*. The precise position of the vortices is obtained from the images after digitally removing scan lines and finding the local minima in $G(V)$ using WSxM software[22]. Each image contains between 250-380 vortices. To identify topological defects, we Delaunay triangulate the VL and determine the nearest neighbor coordination for each flux lines. Topological defects in the hexagonal lattice manifest as points with 5-fold or 7-fold coordination number. The bulk pinning property of the VL is measured on the same crystal from the real part of a. c. susceptibility ($\chi'$) using a home built susceptometer.

**III. Results and Discussion**

We first show the bulk pinning properties (Fig. 1) of the VL at 15 kOe created using different field cooling protocols, measured from the temperature variation of $\chi'$. The two primary states are the zero field cooled (ZFC) state (where the sample is cooled to the lowest temperature in zero field and subsequently a magnetic field is applied) exhibiting weakest pinning and the field cooled (FC) state (where the field is applied at $T > T_c$ and the sample is cooled in the presence of magnetic field) exhibiting strongest pinning. The ZFC state shows a diffuse "peak effect" at $T_p \sim 4.6$ K (Fig. 1(b)), indicating that the sample is moderately disordered[10]. A number of intermediate states with progressively increasing pinning can be accessed by first preparing the VL using the ZFC protocol, and then heating the sample to a higher temperature $T^*(<T_c)$ and cooling back (Fig. 1(a)). We observe that when the FC VL is shaken by applying a small magnetic pulse of 130 Oe (Fig. 1(b)), the VL undergoes a dynamic transition to a state very close to the ZFC state, and after four successive pulses it becomes indistinguishable from the ZFC state. On the other hand, the bulk pinning response of the ZFC state remains unaltered with application of field pulses, which confirms that the ZFC state is the equilibrium state of the system.

Fig. 2(a) and 2(b) show the real space VL image over an area of 500×500 nm, for the ZFC and FC state at 15 kOe and 350 mK. By analyzing the Delaunay triangulation of the structures we observe that the ZFC state is free from topological defects. The vortex lattice constant, $a \approx 39.8$ nm, is in excellent agreement with expected value at 15 kOe. In contrast, we identify three dislocation pairs in the FC lattice within the field of view. When the FC lattice is shaken with a magnetic field pulse of 100 Oe two of the three dislocations are annihilated (Fig. 2(c)). Repeating this experiment with a pulse of 130 Oe we observe that all dislocations are annihilated resulting in state similar to the ZFC state (Fig. 2(d)).

In order to analyze the VL images more quantitatively we calculate the orientational and positional correlation functions, $G_6(\bar{r})$ and $G_{\bar{K}}(\bar{r})$, which measure the degree of misalignment and the relative displacement between two vortices separated by distance $r$ respectively, with respect to the lattice vector of an ideal hexagonal lattice. The orientational correlation function is defined as, $G_6(r) = (1/n(r, \Delta r)) \left\langle \sum_{i,j} \Theta\left(\frac{\Delta r}{2} - \left|r - |\bar{r}_i - \bar{r}_j|\right|\right) \cos 6(\theta(\bar{r}_i) - \theta(\bar{r}_j)) \right\rangle$, where $\Theta(r)$ is the Heaviside step function, $\theta(\bar{r}_i) - \theta(\bar{r}_j)$ is the angle between the bonds located at $\bar{r}_i$ and the bond located at $\bar{r}_j$, $n(r, \Delta r) = \sum_{i,j} \Theta\left(\frac{\Delta r}{2} - \left|r - |\bar{r}_i - \bar{r}_j|\right|\right)$, $\Delta r$ defines a small window of the size of the pixel around $r$ and the sums run over all the bonds. We define the position of each bond as the coordinate of the mid-point of the bond. Similarly, the spatial correlation function, $G_{\bar{K}}(r) = (1/N(r, \Delta r)) \left\langle \sum_{i,j} \Theta\left(\frac{\Delta r}{2} - \left|r - |\bar{R}_i - \bar{R}_j|\right|\right) \cos \bar{K} \cdot (\bar{R}_i - \bar{R}_j) \right\rangle$, where $K$ is the reciprocal lattice vector obtained from the Fourier transform, $R_i$ is the position of the $i$-th vortex, $N(r, \Delta r) = \sum_{i,j} \Theta\left(\frac{\Delta r}{2} - \left|r - |\bar{R}_i - \bar{R}_j|\right|\right)$ and the sum runs over all lattice points. We restrict the range of $r$ to half the lateral size of each image, which corresponds to $7a$ (where $a$ is the average lattice constant) at 15 kOe. For an ideal hexagonal lattice, $G_6(r)$ and $G_{\bar{K}}(r)$ shows sharp peaks with unity amplitude around $1^{st}$, $2^{nd}$, $3^{rd}$ etc... nearest neighbour distance for the bonds and the lattice points respectively. As the lattice disorder increases, the amplitude of the peaks decay with distance and neighbouring peaks at large *r* merge with each other.

Fig. 3 (a) and (b) show $G_6(r)$ and $G_K(r)$ respectively for the states corresponding to Fig. 2(a)-(d). For the ZFC state, we fit $G_K(r)$ with both exponential and power law decay. The exponent for power law fit is -0.03 and the decay constant for exponential fit is very small (0.008) which indicates that $G_K(r) \propto 1/r^n$ (Fig. 3(d)). $G_6(r)$ reach a constant value ~0.85 after 1.5 lattice constant (Fig. 3(c)). This is consistent with long range orientational order and a quasi-long range positional order expected for a BG[23,24]. We confirmed that the intermediate states shown in Fig. 1(a), are similar to the FC state but containing different density of topological defects.

We now address the central issue concerning the glassiness in the BG state. Since a glassy state is expected to contain many degenerate metastable states with same degree of positional order, perturbing the equilibrium BG state, one would expect to switch between these states. To verify this, we prepare the VL in the ZFC state and shake the lattice with a series of 130 Oe magnetic field pulses. While the VL images obtained after applying successive pulses (Fig. 4(a)-(d)) show that they are free from topological defects, after each pulse the VL configuration is different. The difference between successive images (after subtracting the mean value of the conductance over the entire area and normalising the amplitude of the vortex signal to one) is shown in Fig. 4(e)-(g). The difference between the image before and after each pulse shows that after each pulse the VL in some parts of the image drastically rearrange, whereas in other parts it remains relatively unchanged. Fig. 4(h) and (i) show the $G_6(r)$ and $G_K(r)$ respectively corresponding to the images in Fig. 4(a)-(d). The radial decay of $G_6(r)$ and $G_K(r)$ for each of these states is similar and characteristic of a Bragg glass. Thus after each pulse we realise a different metastable realisation of the equilibrium BG state having the same degree of positional and orientational order.

We have independently verified that the lateral drift due to magnetic field pulse in our STM is well below 6.6 nm (see supplementary material), such that the same area is imaged in each of these scans. To verify that the change in position of vortices observed after pulsing is not due to any possible drift in scan area below 6.6 nm, we shift the ZFC lattice by one pixel (6.25 nm) diagonally and take its difference with original ZFC images. We observe that the hexagonal lattice pattern is retained in the image difference, show in Fig. 4(j), whereas the image differences after applying magnetic field pulses show regions of high and low difference indicating rearrangement of vortices.

**IV. Summary**

In summary, we have shown that in the presence of weak random pinning, the equilibrium VL state below the thermodynamic order-disorder transition is quasi-long-range ordered BG, composed of a large number of metastable states, each characterised by the absence of topological disorder and an algebraic decay of the positional correlation. We also show that dislocations in a disordered vortex lattice can be gradually annealed by shaking the lattice with small magnetic field pulses, thus driving it into a BG state. These results are directly relevant to design strategies for growing large self-assembled periodic structures on a substrate[25,26,27], a field of great contemporary interest in material science and photonics[28], where the main challenge is to extend the defect free structure over large length scales. It would be instructive to study the evolution of the BG state as a function of pinning strength, and find the critical value at which the equilibrium state of the system is driven into disordered vortex glass. Such experiments could be performed on colloidal crystals in random optical traps where the strength the pinning potential can be tuned by the strength of optical field, and would provide further insight on the role of random pinning on the crystalline state.

**Acknowledgements**

The authors thank Shobo Bhattacharya, Arun Kumar Grover, Srinivasan Ramakrishnan and Deepak Dhar for illuminating discussions during the course of the work. The work was funded by Department of Atomic Energy, Government of India. PS acknowledges Department of Science and Technology, Government of India for partial financial support through grant no SR/S2/RF-121/2012.

[1] M. J. Higgins and S. Bhattacharya, Physica C **257**, 232(1996).

[2] G. Blatter, M. V. Feigel'man, V. B. Geshkenbein, A. I. Larkin and V. M. Vinokur, Rev. Mod. Phys. **66**, 1125 (1994).

[3] Y. Paltiel, E. Zeldov, Y. N. Myasoedov, H. Shtrikman, S. Bhattacharya, M. J. Higgins, Z. L. Xiao, E. Y. Andrei, P. L. Gammel and D. J. Bishop, Nature **403**, 398 (2000).

[4] T. Giamarchi and P. Le Doussal, Phys. Rev. B **52,** 1242 (1995).

[5] T. Klein, I. Joumard, S. Blanchard, J. Markus, R. Cubitt, T. Giamarchi and P. Le Dousal, Nature **413,** 404 (2001).

[6] D. S.Fisher, M. P. A.Fisher and D. A. Huse, Phys. Rev. B **43**, 130 (1991).

[7] W. Henderson, E. Y. Andrei, M. J. Higgins and S. Bhattacharya, Phys. Rev. Lett. **77,** 2077 (1996).

[8] S. Mohan, J. Sinha, S. S. Banerjee, A. K. Sood, S. Ramakrishnan and A. K. Grover, Phys. Rev. Lett. **103**, 167001 (2009).

[9] K. Ghosh, S. Ramakrishnan, A. K. Grover, G. I. Menon, G. Chandra, T. V. Chandrasekhar Rao, G. Ravikumar, P. K. Mishra, V. C. Sahni, C. V. Tomy, G. Balakrishnan, D.Mck Paul and S. Bhattacharya, Phys. Rev. Lett. **76**, 4600 (1996).

[10] S. S. Banerjee, N. G. Patil, S. Ramakrishnan, A. K. Grover, S. Bhattacharya, G. Ravikumar, P. K. Mishra, T. V. Chandrasekhar Rao, V. C. Sahni and M. J. Higgins, Appl. Phys. Lett. **74** , 126 (1999).

[11] G. Pasquini, D. Pérez Daroca, C. Chiliotte, G. S. Lozano and V. Bekeris, Phys. Rev. Lett. **100**, 247003 (2008).

[12] G. Ravikumar, V. C. Sahni, A. K. Grover, S. Ramakrishnan, P. L. Gammel, D. J. Bishop, E. Bucher, M. J. Higgins and S. Bhattacharya, Phys. Rev. B **63**, 024505 ( 2000).

[13] H. Pastoriza, M. F. Goffman, A. Arribére and F. de la Cruz, Phys. Rev. Lett. **72**, 2951 (1994).

[14] A. Larkin and Y. Ovchinnikov, J. Low Temp. Phys. 34, 409 (1979).

[15] A. M. Troyanovski, M.van Hecke, N. Saha, J. Aarts and P. H. Kes, Phys. Rev. Lett. **89**, 147006 (2002).

[16] A. P. Petrović, Y. Fasano, R. Lortz, C. Senatore, A. Demuer, A. B. Antunes, A. Paré, D. Salloum, P. Gougeon, M. Potel and Ø. Fischer, Phys. Rev. Lett. **103**, 257001 (2009).

[17] I. Guillamón, H. Suderow, A. Fernández-Pacheco, J. Sesé, R. Córdoba, J. M. De Teresa, M. R. Ibarra and S.Vieira, Nature Physics **5**, 651 (2009).

[18] S. S.Banerjee, N. G.Patil, S. Ramakrishnan, A. K. Grover, S. Bhattacharya, P. K. Mishra, G. Ravikumar, T. V. Chandrasekhar Rao, V. C. Sahni, M. J. Higgins, C. V. Tomy, G. Balakrishnan and D. Mck. Paul, Phys. Rev. B **59**, 6043 (1999).

[19] A. D. Thakur, T. V. Chandrasekhar Rao, S.Uji, T. Terashima, M. J. Higgins, S. Ramakrishnan and A. K. Grover, J. Phys. Soc. Jpn. **75**, 074718 (2006).

[20] See Supplemental Material at [*URL will be inserted by publisher*] for thermodynamic order-disorder transition driven by magnetic field at 350 mK.

[21] A. Kamlapure, G. Saraswat, S. C. Ganguli, V. Bagwe, P. Raychaudhuri and S. P. Pai, Rev. Sci. Instrum. **84**, 123905 (2013).

[22] I. Horcas, R. Fernández, J. M.Gómez-Rodríguez, J. Colchero, J. Gómez-Herrero and A. M. Baro, Rev. Sci. Instrum. **78**, 013705 (2007).

[23] T. Nattermann, Phys. Rev. Lett. **64**, 2454 (1990).

[24] T. Giamarchi and P. le Doussal, Phys. Rev. Lett **72**, 1530 (1994).

[25] A. M.Alsayed, M. F. Islam, J. Zhang, P. J. Collings and A. G. Yodh, Science **309**, 1207 (2005).

[26] P. Pieranski, L. Strzelecki and B. Pansu, Phys. Rev. Lett. **50,** 900 (1983).

[27] T. Zhang, X. Tuo and J. Yuan, Langmuir **25,** 820 (2009).

[28] J. F. Galisteo-López, M. Ibisate, R. Sapienza, L. S. Froufe-Pérez, Á. Blanco and C. López, Adv. Mater. **23,** 30 (2011).

**Figure Captions:**

**Figure 1.** (a) $\chi'$ vs. T at 15 kOe for the VL prepared using different protocols. (b) $\chi'$ vs. T for the ZFC and FC states and after applying different number of magnetic field pulses of 130 Oe on the FC state. $\chi'$ is normalised to zero field value at lowest temperature. The data is recorded while heating the sample.

**Figure 2.** Real space image (500× 500 nm) of the VL at 350 mK in magnetic field of 15 kOe after preparing the VL using different protocols: (a) ZFC, (b) FC, (c) after applying a pulse of 100 Oe and (d) after applying a pulse of130 Oe on the FC state. The lines show Delaunay triangulation and the dislocations in the VL are shown as pairs of points with five-fold (red) and seven-fold (white) coordination. Corresponding Fourier transforms are shown on the right panel of each real space image.

**Figure 3.**(a) Orientational correlation function,$G_6$ and (b) and positional correlation function, $G_K$ (averaged over the principal symmetry directions) as a function of *r/a* for the VL configurations shown in Fig. 2(a)-(d). *a* is calculated by averaging over all the bonds after Delaunay triangulating the image. (c) is an expanded view showing that $G_6$ for ZFC is constant ~ 0.85 (red line) after 2 lattice constants. Panel (d) shows the fit to $G_K$ for ZFC by exponential/power-law decay respectively. The exponent for power law fit is -0.03 and the decay constant for exponential fit is 0.008 which indicates that the decay is a power law.

**Figure 4.** Real space image (670×670 nm) at 350 mK of (a) the ZFC VL at 15 kOe and (b)-(d) after applying successive magnetic field pulses of 130 Oe. Delaunay triangulations of the VL are shown in the same figures. After each pulse the VL configuration is altered which is apparent from the difference between images before and after each pulse: (e) (b)-(a) (f) (c)-

(b) and (g) (d)-(c). (h) Orientational correlation function, $G_6$ and (i) positional correlation function, $G_K$ (averaged over the principal symmetry directions) as a function of *r/a* for the VL configurations shown in panels(a)-(d). We confirm that the drift in scan area is less than 6.6 nm (Supplementary material). By shifting the ZFC VL configuration diagonally by 1 pixel = 6.25 nm, and taking the image diffrence between ZFC and shifted ZFC, we observe that we retain the hexagonal lattice pattern in the image difference. Whereas the image differences after applying magnetic field pulses show regions on high and low difference indicating that it is not an artifact due to drift in the image.

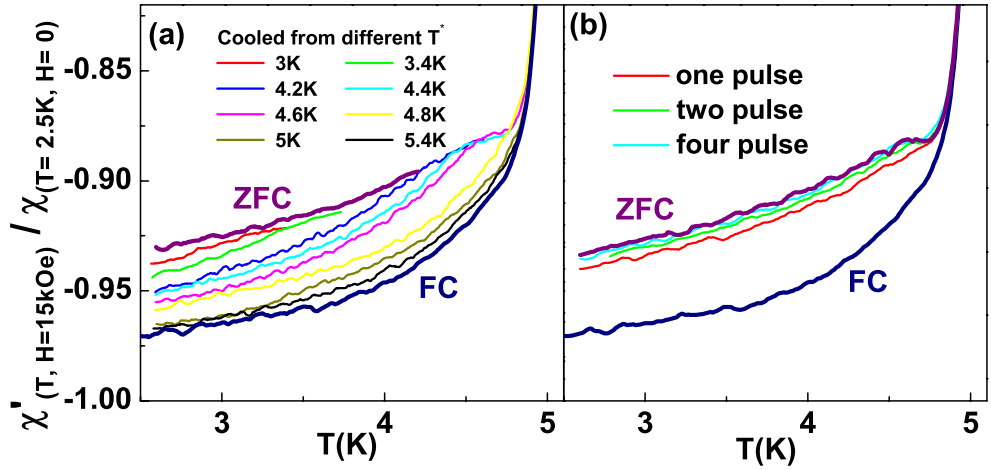

**Figure 1.** (a) $\chi'$ *vs.* $T$ at 15 kOe for the VL prepared using different protocols. (b) $\chi'$ *vs.* $T$ for the ZFC and FC states and after applying different number of magnetic field pulses of 130 Oe on the FC state. $\chi'$ is normalised to zero field value at lowest temperature. The data is recorded while heating the sample.

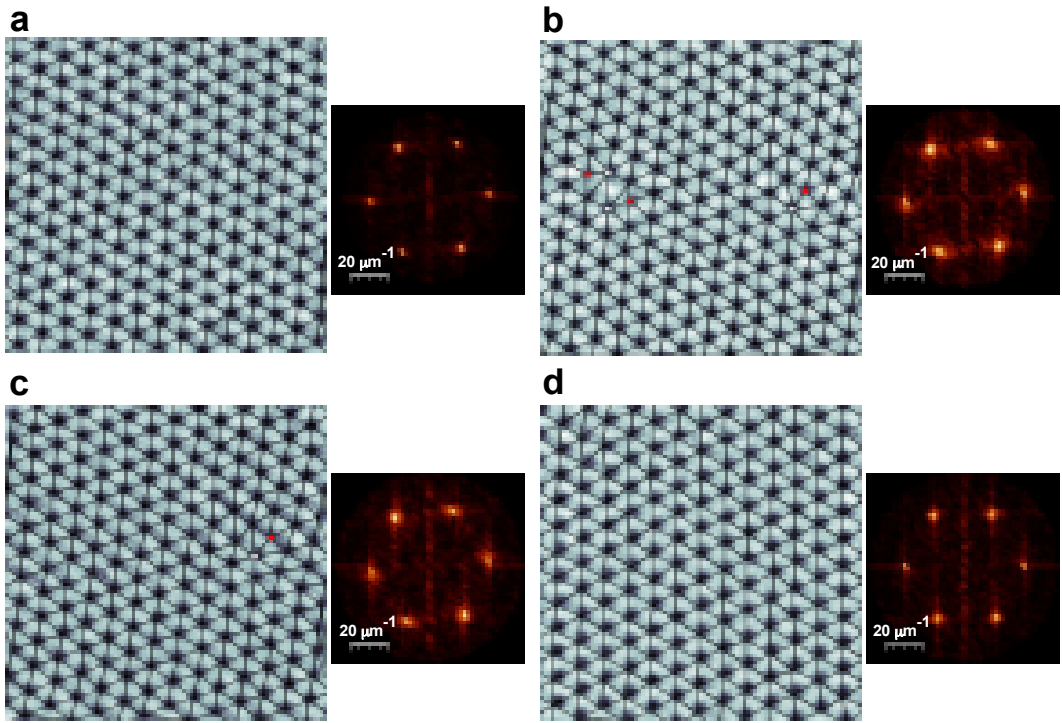

**Figure 2.** Real space image (500× 500 nm) of the VL at 350 mK in magnetic field of 15 kOe after preparing the VL using different protocols: (a) ZFC, (b) FC, (c) after applying a pulse of 100 Oe and (d) after applying a pulse of 130 Oe on the FC state. The lines show Delaunay triangulation and the dislocations in the VL are shown as pairs of points with five-fold (red) and seven-fold (white) coordination. Corresponding Fourier transforms are shown on the right panel of each real space image.

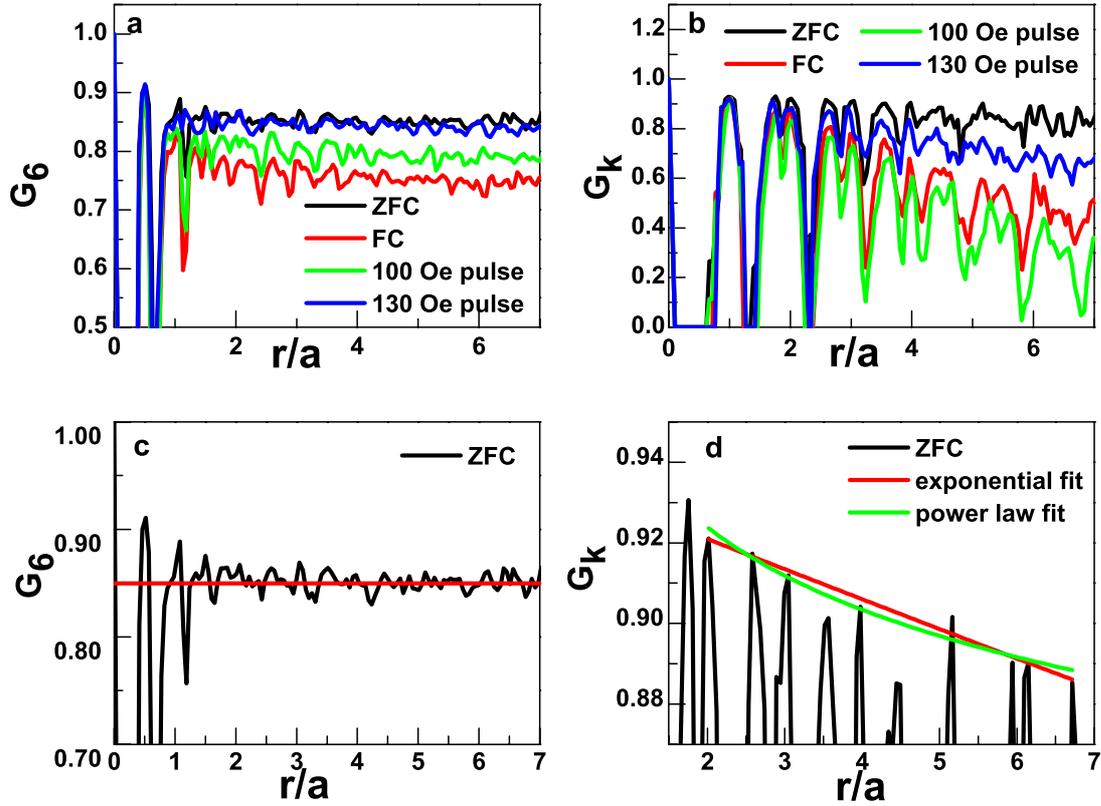

**Figure 3.** (a) Orientational correlation function, $G_6$ and (b) and positional correlation function, $G_K$ (averaged over the principal symmetry directions) as a function of $r/a$ for the VL configurations shown in Fig. 2(a)-(d). $a$ is calculated by averaging over all the bonds after Delaunay triangulating the image. (c) is an expanded view showing that $G_6$ for ZFC is constant ~ 0.85 (red line) after 2 lattice constants. Panel (d) shows the fit to $G_K$ for ZFC by exponential/power-law decay respectively. The exponent for power law fit is -0.03 and the decay constant for exponential fit is 0.008 which indicates that the decay is a power law.

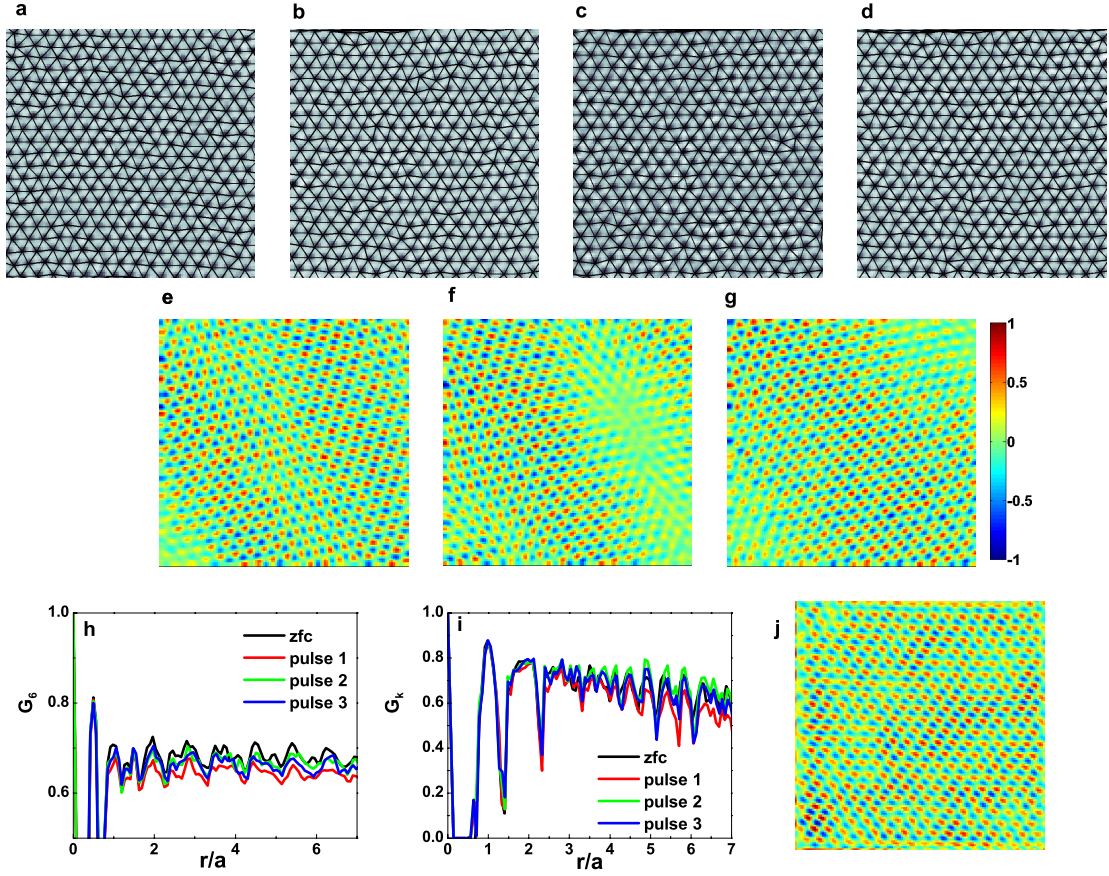

**Figure 4.** Real space image (670×670 nm) at 350 mK of (a) the ZFC VL at 15 kOe and (b)-(d) after applying successive magnetic field pulses of 130 Oe. Delaunay triangulations of the VL are shown in the same figures. After each pulse the VL configuration is altered which is apparent from the difference between images before and after each pulse: (e) (b)-(a) (f) (c)-(b) and (g) (d)-(c). (h) Orientational correlation function, $G_6$ and (i) positional correlation function, $G_K$ (averaged over the principal symmetry directions) as a function of $r/a$ for the VL configurations shown in panels(a)-(d). We confirm that the drift in scan area is less than 6.66 nm (Supplementary material). By shifting the ZFC VL configuration diagonally by 1 pixel = 6.25 nm, and taking the image diffrence between ZFC and shifted ZFC, we observe that we retain the hexagonal lattice pattern in the image difference. Whereas the image differences after applying magnetic field pulses show regions on high and low difference indicating that it is not an artifact due to drift in the image.

*Supplementary Material*

**S1. Thermodynamic order-disorder transition driven by magnetic field at 350 mK**

The experiments reported in this paper are carried out at fields and temperature much below the locus of the thermodynamic order disorder transition. The thermodynamic disordering field at 350 mK is independently estimated by preparing the VL in the ZFC state and recording the VL image with increasing magnetic field (Fig. 1s). We observe that till 2.8 T the VL is free from topological defects. The onset of the disordering of the VL is seen at 3.0 T through spontaneous proliferation of dislocations. Consequently, all our measurements are performed at 1.5 T, well below the disordering field.

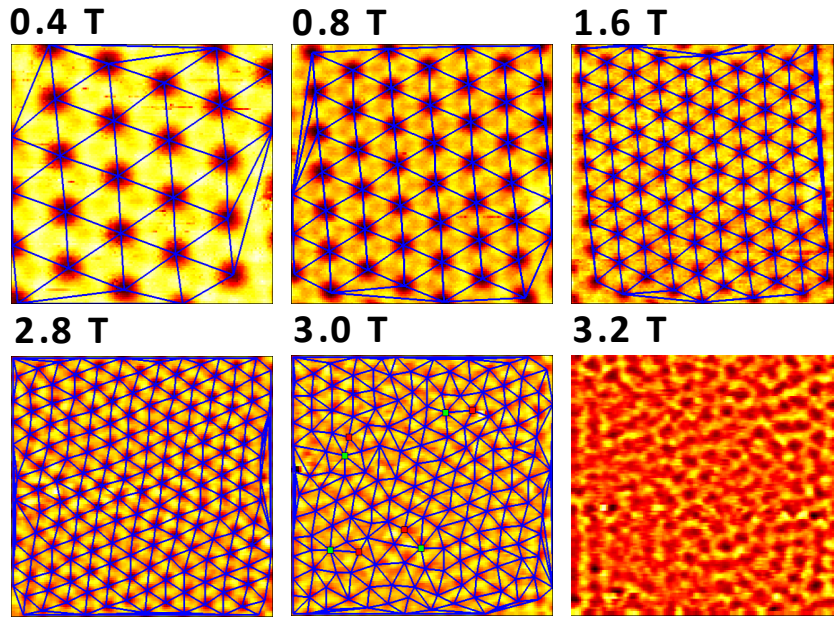

**Figure 1s.** VL images (352 × 352 nm) at different fields at 350 mK taken while increasing the magnetic field after zero field cooling the sample. The Delaunay triangulation of the VLs (blue lines) shows the appearance of dislocations at 3 T. The red and green points correspond to sites with 5-fold and 7-seven fold coordination. The contrast at 3.2 T is too poor to reliably locate the position of individual vortices.

## S2. Drift in scan area by applying a magnetic field pulse

The drift in scan area due to application of a 100 Oe pulse is estimated by identifying topological features in the scanned area before and after applying pulse. In 40 pixel by 40 pixel resolution image of a 266 x 266 nm$^2$ area, each pixel corresponding to 6.6nm. There is no percievable drift at imaged resolution. Thus we conclude that the drift caused by application of magnetic field pulse is ≤ 6.6 nm in any direction.

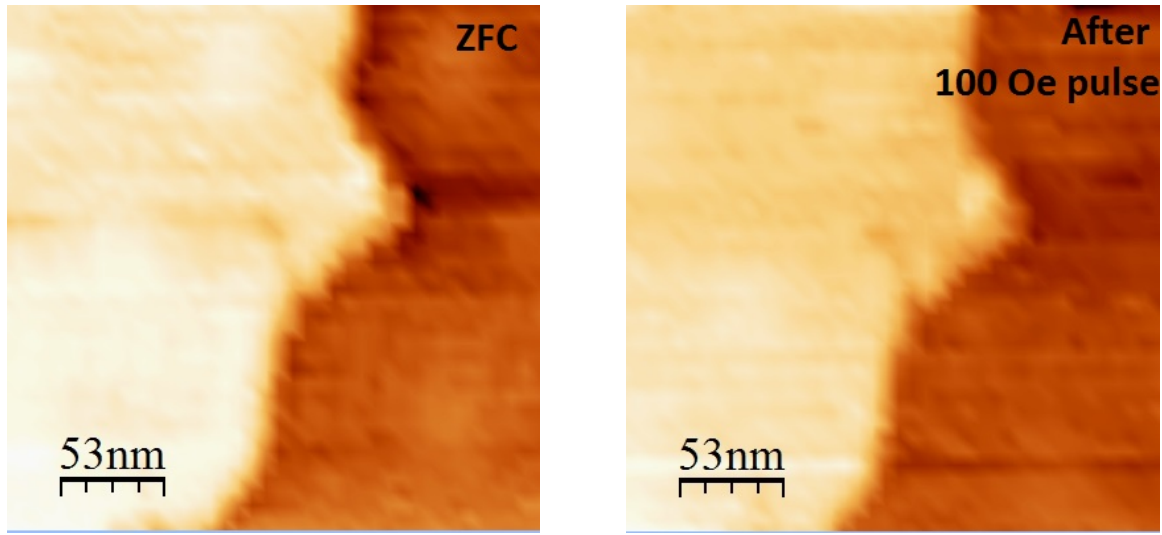

**Figure 2s.** Topographic images (266 × 266 nm$^2$) taken first after preparing a ZFC state and then after applying 100 Oe pulse. The image on the right, after applying the magnetic pulse shows no percievable drift thus we conclude the drift to be less than one pixel ~ 6.6 nm.